\documentclass[12pt,preprint]{aastex}

\shorttitle{Proper motion of PSR 1929+10} 
\shortauthors{Mignani et al.}
\begin{document}

\title{HST Proper Motion confirms the optical identification of the nearby pulsar PSR 1929+10\footnote{Based
on observations with the NASA/ESA Hubble Space Telescope,
obtained at the Space Telescope Science Institute, which is
operated by AURA, Inc.  under contract No NAS 5-26555}}

\author{Roberto P.~Mignani}
\affil{European Southern Observatory, Karl Schwarzschild Str. 2,
D-85740, Garching, Germany}
\email{rmignani@eso.org}
\author{Andrea De Luca,Patrizia A. Caraveo}
\affil{Istituto di Astrofisica Spaziale e Fisica Cosmica, Sezione di
Milano "G.Occhialini" - CNR v. Bassini 15, I-20133 Milan, Italy}
\email{deluca@mi.iasf.cnr.it,pat@mi.iasf.cnr.it}
\author{Werner Becker}
\affil{Max Planck Institute f\"ur Extraterrestrische Physik, D-85748, Garching, Germany }
\email{web@mpe.mpg.de}

\begin{abstract}
We  report on the proper  motion  measurement of  the proposed optical
counterpart of the X-ray/radio   pulsar  PSR 1929+10.  Using    images
obtained with the   HST/STIS (average epoch   2001.73)  we computed  a
yearly displacement of $+97 \pm 1$ mas yr$^{-1}$ in RA and $+46 \pm 1$
mas yr$^{-1}$ in Dec since the epoch (1994.52) of the original HST/FOC
detection.  Both   the magnitude and  direction of  the optical proper
motion components are   found to be  fully  consistent with  the  most
recent VLBA radio measurements.   This result provides an  unambiguous
confirmation of  the pulsar optical  identification.  In  addition, we
have used the combined STIS/FOC datasets  to derive information on the
pulsar spectrum,  which seems characterized by  a power  law component, apparently unrelated to the X-ray emission. 
\end{abstract}


\section{Introduction}

PSR1929+10 is an old  ($ \sim 3~  10^{6}$ yrs)  radio pulsar.   With a
distance  of  $\sim$   330  pc,  determined from   VLBA radio parallax
measurements (Brisken et al.  2002), it is also one  of the closest to
the solar  system.  After the  original  X-ray detection with Einstein
(Helfand   1983),  pulsations at the    radio   period (227 ms)   were
discovered by ROSAT  (Yancopulous et al. 1994)  and later confirmed in
ASCA data (Wang and Halpern  1997).  The X-rays pulse profile exhibits
a single, broad,  peak markedly  different  from the sharp  radio one.
The X-ray spectrum can be described either by a blackbody ($T \sim 3-5
~ 10^{6}~K$) produced  from hot polar caps  (Yancopulous et al.  1994;
Wang and Halpern 1997) or by a power law with $\alpha \approx 1.27 \pm
0.4$ (Becker  and Tr\"umper 1997).  A trail  of diffuse X-ray emission
originated from the  pulsar and extending $\sim$   10 arcmin to  South
East    was discovered with  the ROSAT/PSPC   (Wang  et al.  1993).  A
candidate  optical counterpart to PSR1929+10  was  identified with the
HST/FOC by Pavlov et  al.  (1996) based  on the positional coincidence
($\sim 0\farcs4$)   with  the radio  coordinates.  Interestingly,  the
measured flux of the PSR1929+10 counterpart  ($U \sim 25.7$) was found
to deviate by 3 orders of magnitude from the values predicted from the
X-ray spectra.  This behaviour is  markedly different from that of the
middle aged pulsars PSR0656+14 (Pavlov et al.  1997), Geminga (Mignani
et al.  1998) and PSR1055-52 (Mignani et  al. 1997), where the optical
data are not  too  far from the   extrapolation of the  X ray spectra.
Confirming the optical identification  of PSR1929+10 becomes a crucial
step to settle  a consistent scenario for the  long term evolution  of
the   optical  luminosity  of  pulsars  and  to  investigate  possible
turnovers in the emission physics.  While young ($ \sim 10^{3}-10^{4}$
yrs) objects are relatively bright, their  optical throughput seems to
decay on a timescale of few thousands  years and progressively turn to
a composite magnetospheric/thermal   regime  (Mignani   1998;  Caraveo
2000).  Although  evidence  for  such  a trend  can be  recognized  in
middle-aged objects ($\sim 10^{5}$ yrs), like PSR 0656+14 and Geminga,
little is known on the optical emission at later  stages of the pulsar
lifetime. \\  Taking advantage of   new HST observations, we  use  the
pulsar proper motion to secure the PSR 1929+10 optical identification,
thus adding an important piece of information on the optical behaviour
of old pulsars.  Observations and data reduction are described in \S2,
while the results are discussed in \S3.

\section{Observations}

 The field of  PSR 1929+10 was  observed with the STIS detector aboard
 HST in five different  visits on  August  28th 2001, September  11th,
 15th, 21st 2001, October 20th  2001.  The observations were performed
 after the reactivation of the STIS  following problems with the power
 supply  units occurred in  May 2001.  Owing  to this malfunction, the
 observations  could have   been affected  by  a residual  few percent
 increase of the   dark current  level.    For each  visit, the  total
 integration time  was 2\,400  s, split  in two exposures  of 1\,200 s
 each.  Observations run smoothly apart from short gaps ($\sim $ 60 s)
 in the engineering telemetry occurred during  visits \#3 and \#4. The
 NUV-MAMA  detector,  with a  pixel size   of $0\farcs024$ ($24\farcs7
 \times  24\farcs7$  field of view),  was  used  its  TIME-TAG mode to
 obtain time-resolved  images  with a  temporal   resolution of  $ 125
 \mu$s.   To add  spectral  information in the   NUV and to complement
 previous data obtained with the FOC  at $ \sim  3\,400 \AA$ by Pavlov
 et al.  (1996), the  exposures were taken  through the F25QTZ  filter
 ($\lambda=2364  \AA,  \Delta \lambda  \sim   842 ~\AA~FWHM$).  \\ Our
 identification  strategy   was   based    on two  independent     and
 complementary  approaches.  Firstly,   following the straightforward,
 but powerful,  strategy   successfully applied  in   the case  of the
 identification of PSR 0656+14 (Mignani, De Luca and Caraveo 2000), we
 performed a proper motion  measurement of the proposed counterpart to
 be    compared with the known   one   of the  radio pulsar,  recently
 reassessed  by Brisken et al.   (2002) using  the VLBA.  Secondly, we
 used our  time-resolved images to search  for pulsations at the radio
 period from the candidate counterpart.  The proper motion measurement
 is  described  in the following   sections, while the  results of the
 timing  analysis are presented in a  companion paper  (Mignani et al.
 in preparation).

\subsection{Data Analysis}

As starting point  for our proper motion measurement  we used  the FOC
observations of  Pavlov et al.   (1996), collected on  July 10th 1994.
Images were taken in  three different filters: F130LP ($\lambda=3437.7
\AA, \Delta \lambda  \sim 1965 ~\AA~FWHM$), F342W ($\lambda=3402  \AA,
\Delta  \lambda \sim  442  ~\AA~FWHM$)  and F430W ($\lambda=3940  \AA,
\Delta \lambda \sim  832~\AA~FWHM$),  with total integration  times of
1221,  3310 and 596  s, respectively.  The camera  was operated at two
different focal lenghts corresponding to a  field of view of $7\farcs4
\times 7\farcs4$ for the F130LP and F342W  exposures and of $14\farcs8
\times 14\farcs8$  for  the F430W  one.   In both cases,  the  angular
resolution was $0\farcs014$  per pixel.  The  data were retrieved from
the ST-ECF public archive after on-the-fly recalibration with the best
reference files.  As  shown  by  Pavlov et  al.   (1996), the   pulsar
counterpart is detected only through the  F130LP and the F342W filters
(see their figures 2c and 3a).  \\ The STIS images were retrieved from
the STScI  data archive after the default  pipeline calibration.  Each
image has been corrected for the geometric distortion of the CCD using
the     DRIZZLE    software     distributed      in   the       STSDAS
package\footnote{stsdas.stsci.edu/STSDAS.html} and applying  the  most
recent coefficients of the cubic mapping of the NUV-MAMA field of view
listed         in              the               STIS           Users'
Handbook\footnote{www.stsci.edu/hst/stis/documents/handbooks/cycle11/stis\_cy11\_ihbTOC.html}.
The  two  exposures  taken    during each visit  were   coadded  after
accounting for thiny  shifts (of order  0.4 pixels) and differences in
roll  angle (of order  0.03$^{\circ}$)  due to  the spacecraft jitter.
For each couple,   the average coadditon  accuracy  was of  $\sim$ 0.1
pixels in both $x$ and $y$. 

Since  the expected  overall  displacement of  the pulsar  counterpart
($\sim$ 15  mas, equivalent to 0.6  STIS pixels) in the epoch interval
spanned by our observations (2001.65-2001.80) would be negligible with
respect to the total displacement ($\sim$ 770 mas) predicted since the
reference  epoch (1994.52), we  first combined all  the five available
STIS images to benefit of the higher $S/N$.  Using as relative reference
grid the  coordinates  of 7   to 10 common   stars  (the actual number
depending on the telescope roll angle), all the frames were registered
on the mid-epoch  frame  (i.e. the september  15th  one) by fitting  a
linear coordinate transformation after  aligning  each frame in  right
ascension and declination according to the  telescope roll angles. The
final  STIS image resulting from the  combination of all the available
frames  is shown in Figure  1.   As expected in  the  case of a moving
object, the centering  accuracy of the   target in the  combined image
(0.3 pixels in RA and 0.15 pixels in Dec)  appears degraded wrt to the
values  measured  in  each single  single frame,   for which we  found
accuracies of $\sim$ 0.1 pixels in both coordinates.   To the error on
the centroid determination we then added in  quadrature the rms on the
epoch-to-epoch  coordinate  transformation, which in all  cases turned
out to be within 0.2 pixels (per coordinate), plus the accuracy of the
exposures coaddition in each visit  (0.1 pixels).  The final precision
on the target position in the combined STIS image was thus 0.37 pixels
in RA and 0.27 pixels in Dec. 

Next  step was to apply the  STIS-to-FOC  registration to evaluate the
relative displacement  of the pulsar counterpart  over  the $\sim$ 7.2
years  interval spanned by   the available observations.  Although our
target was clearly detected in the  FOC/F130LP image (preferred to the
F342W one  because of its higher $S/N$),  its $7\farcs4 \times 7\farcs4$
field of view containes only one object in  common with the wider STIS
image ($24\farcs7 \times  24\farcs7$).  Thus,  we decided to  register
both images on a common reference frame defined by the FOC/F430W image
($14\farcs8 \times  14\farcs8$).  This allowed  us to use three bright
reference objects     for the    FOC/130LP-to-FOC/430W   superposition
(accounting for shift and rotation for a resulting accuracy of 0.3 FOC
pixels   per  coordinate) and  4   good   reference  objects  for  the
STIS-to-FOC/430W superposition (rms of 0.55  FOC pixels per coordinate
- accounting for  shift,   rotation and  scale factor).   The  overall
accuracy in   the  final STIS-to-FOC/F130LP   superposition  was  thus
obtained  by adding in  quadrature  all the uncertainties quoted above
and turned out to be $\sim$ 0.6 FOC pixels per coordinate. 

\section{Results}

\subsection{Proper Motion}

After registering both the   FOC  and the   STIS images on  a   unique
reference  frame, we  could evaluate the  displacement   of the pulsar
candidate  counterpart by simply   measuring   the difference in   its
relative coordinates.  The overall  uncertainty on such difference was
estimated by  adding in quadrature  all the uncertainties derived from
the   different steps  of  our  relative  astrometry procedure.  These
include:  the  centering  error  of the  counterpart   in the combined
five-epochs STIS image (0.37 STIS pixels in RA and 0.27 pixels in Dec)
and in the  FOC/130LP image ($\sim$  0.1 FOC pixels  per coordinate)
plus the overall  accuracy of the STIS-to-FOC/130LP superposition (0.6
FOC  pixels per coordinate)  as  obtained  by  the two-step  procedure
described in  the previous  section.   The difference in the  relative
coordinates is 48.7 $\pm$ 1.0 FOC  pixels and 23.2 $\pm$
1.0 FOC  pixels along the RA  and Dec  directions, respectively.  Such
measurements  represent a clear
evidence for  the object displacement over  the $\sim$  7.2 years time
span between the FOC (1994.52)  and the average STIS epoch (2001.731).
The pulsar displacement  can  be appreciated  in  Figure 2, where  the
relative FOC position is overlayed  on the STIS image.  After applying
the FOC plate  scale of $0\farcs01435$/pixel (with  0.5\% uncertainty)
to translate from  pixel  to sky  coordinates, we  computed the  proper
motion of the pulsar optical counterpart. This is $\mu_{\alpha}cos(\delta) =
+97 \pm 2$ mas yr$^{-1}$ and $\mu_{\delta} = +46 \pm 2$ mas
yr$^{-1}$, corresponding to a total yearly displacement in the plane of
the  sky of $\mu =  107.3 \pm 1$ mas yr$^{-1}$  along a position angle
(PA) of 64.6$^{\circ}$ $\pm$ 0.5$^{\circ}$.  Although the FOC and STIS
observations were  taken at  different  times of the  year,  we did not
apply any correction for  the object's parallax  (Brisken et al. 2002)
as the effect of the tiny parallactic displacement  is well within our
error budget. 

As  a further check, we have  recomputed the pulsar displacement using
the FOC/130LP image and,  in turn, each of  the five single-epoch STIS
images.  This allowed us to obtain five  independent measures for five
different epoch pairs (see Table 1) and thus to exclude the effects of
unknown systematics in  our procedure.  In   this case, for  each STIS
image the error on the target position was only due to the combination
of   the centering error ($\sim$  0.1  pixels per coordinates) and the
accuracy of the coaddition of the single exposures  in each visit (0.1
pixels).  The final  precision on the target  position in each of  the
five single-epoch STIS images was thus of the order of 0.15 pixels per
coordinate.  The  strategy for  the STIS-to-FOC/130LP registration was
clearly the same  described  above.  We computed  the STIS-to-FOC/430W
frame registrations  (rms of  0.45 -0.65  FOC pixels per  coordinate -
accounting for shift, rotation and scale  factor) with the same number
of  objects used in   \S  2.1.  The  overall   accuracy of the   final
STIS-to-FOC/130LP registrations turned out  to be between 0.55 and 0.8
FOC    pixels.  As described above,   the   uncertainty on the  target
displacement is due  to the combination of the  centering error of the
counterpart  in the single STIS images   (0.15 original STIS pixels in
both  RA and Dec)   and in the FOC/130LP  image   (0.1 FOC  pixels per
coordinate) plus the  overall STIS-to-FOC/130LP superposition accuracy
(0.55-0.8 FOC  pixels per coordinate).   \\ The proper motions derived
from the  displacements computed for each of  the five epoch pairs are
listed in Table 1, where all the values are seen to be consistent with
the proper motion  obtained using the  combined STIS image. A $\chi^2$
fit to the   values listed in Table  1  yields the  best proper motion
values:  $\mu_{\alpha}cos(\delta)    =  +97  \pm 1$   mas~yr$^{-1}$ and
$\mu_{\delta} =  +46 \pm 1$ mas~yr$^{-1}$,  corresponding to  a yearly
displacement  $\mu =  107.35 \pm 1$  mas yr$^{-1}$ along  a position
angle of 64.63$^{\circ}$ $\pm$  0.55$^{\circ}$.  \\ We can now compare
our  best   proper motion value   with the  most recent  radio measure
obtained by Brisken et al.   (2002), who give $\mu_{\alpha}cos(\delta)
= +94.82 \pm 0.26$ mas~yr$^{-1}$ and $\mu_{\delta}  = +43.04 \pm 0.15$
mas~yr$^{-1}$ (65.58$^{\circ}$ $\pm$ 0.09$^{\circ}$ PA).  Although our
value is somewhat less precise, we note that the optical proper motion
is fully consistent  with the  radio  one.   Thus, our proper   motion
measurement provides  a  robust proof  that the  candidate proposed by
Pavlov et   al.  (1996) is  indeed   the  optical counterpart  of  PSR
1929+10.

\subsection{Photometry}

We have used our STIS images to measure  the pulsar flux in the F25QTZ
filter.  In    order to  take advantage    of  the higher   $S/N$, our
photometry  has been computed  on the combined  image (Figure 1).  The
source counts were extracted using an  optimized aperture and the flux
conversion was applied using the photometric zeropoint provided by the
pipeline STIS flux calibration.  We thus  derived a flux of (5.4 $\pm$
0.4)~ $10^{-31}$ erg cm$^{-2}$ s$^{-1}$ Hz$^{-1}$, corresponding to an
ST-mag of 25.4 $\pm 0.15$.  The attached error takes into account both
the accuracy of our aperture photometry and the systematic uncertainty
of  $\sim$  5\% which  affects the absolute  flux  calibration  of the
STIS/MAMA\footnote{www.stsci.edu/hst/stis/documents/handbooks/cycle11/stis\_cy11\_ihbTOC.html}. The
measured flux was then compared with  the values obtained with the FOC
in the 130LP and 342W filters.  For consistency, we have independently
reanalized the FOC datasets finding flux values virtually identical to
the  ones reported in Pavlov et  al.  (1996).   This confirms that the
optical points are definetely inconsistent  with the extrapolations of
the X-ray spectra which, depending on the  assumed model (power law or
polar caps),  predict fluxes $\sim$ 3  orders  of magnitude higher and
lower,  respectively.   The comparison  between the   FOC and the STIS
photometry is shown in  Figure 3, where in  the upper panel we plotted
the measured values  and in the lower panel  we corrected for a likely
upper  limit on the  extinction of   $E(B-V)=0.1$  (see Pavlov et  al.
1996).  While   the data can not  be  accounted by a  single blackbody
function, a  power law provides a more  straightforward fit.   For the
two  extreme values  of the extinction  the  power  law spectral index
$\alpha$ varies between 1    and 0, respectively.  Although  the  data
could  also  be  compatible   with  a composite  model,  any realistic
spectral fit is hampered both by  the few points  available and by the
limited spectral coverage. 

\section{Conclusions}

Using images collected  with the STIS camera  aboard HST together with
archived HST/FOC images taken 7.2 years apart we  have measured a very
significant angular displacement  of the proposed  optical counterpart
tof PSR 1929+10.  This yields a proper motion $\mu_{\alpha}cos(\delta)
=  +97  \pm    1$  mas~yr$^{-1}$  and  $\mu_{\delta}  =    +46 \pm  1$
mas~yr$^{-1}$. These  values  agree with the  ones  derived  from very
recent VLBA radio measurements  (Brisken et al.  2002), thus providing
an unambiguous  confirmation  of the pulsar  identification.  Securing
the   identification of an   old pulsar  such  as   PSR 1929+10  is an
important step to assess the  pulsars' optical behaviour as a function
of their  age.  At   variance  with the phenomenology  of  middle aged
objects, such as PSR 0656+14, Geminga and PSR 1055-52, the optical
emission of PSR 1929+10 seems unrelated to the X-ray one, be it either
of thermal pr non-thermal origin. Although the new STIS data seem to
favour a power law rather than a blackbody, the paucity of flux values
and the limited  spectral coverage available  do not  allow us to  put
firmer constraints  on the spectrum.  More  data, expecially at longer
wavelengths, are required  to better characterize the  pulsar spectral
shape and  to   unveil the  possible presence of    different spectral
components.  The detection of the optical timing signature will add an
important piece of information to understand  the optical behaviour of
this old pulsar. 

\acknowledgements  RM warmly  thanks F.   Patat  for his  help in  the
finalization of this observing program.

\begin{table}
\begin{center}
\begin{tabular}{|l|c|c|c|c|c|c|} \tableline \tableline
Pair & Epoch \#1 & Epoch \#2 & $\mu_{\alpha}cos(\delta)$ & $\mu_(\delta)$ & $\mu$ & PA  \\
\tableline \tableline
1 & 1994.52 & 2001.654 & 97.0$\pm$1.4 & 46.0$\pm$1.4& 107.3$\pm$1.4 &64.63$\pm$0.75\\
2 & 1994.52 & 2001.695 & 97.7$\pm$1.5 & 44.8$\pm$1.5& 107.5$\pm$1.5 &65.36$\pm$0.80\\  
3 & 1994.52 & 2001.704 & 96.6$\pm$1.6 & 48.6$\pm$1.6& 108.1$\pm$1.6 &63.29$\pm$0.84\\
4 & 1994.52 & 2001.720 & 96.7$\pm$1.3 & 45.8$\pm$1.3& 107.0$\pm$1.3 &64.65$\pm$0.70\\
5 & 1994.52 & 2001.808 & 96.8$\pm$1.3 & 45.5$\pm$1.3& 106.9$\pm$1.3 &64.87$\pm$0.69\\

 \tableline \tableline
\end{tabular}
\end{center}                                                                                                 
\caption{Summary of  the proper motion measures  obtained by comparing
the  original pulsar  position  derived from  the  FOC observation  of
Pavlov  et  al.  (1996)  with  those derived  from  each  of the  five
available  STIS observations (see  text). Column  1 numbers  the epoch
pairs, column 2 gives the reference epoch of the FOC observation while
column 3 gives the epochs of the STIS observations. The derived proper
motion values (mas/yr) in RA and Dec, the total proper motion and the position
angle (degrees) are listed in columns 4 to 7, respectively. }
\end{table}

\clearpage

\begin{figure}
\plotone{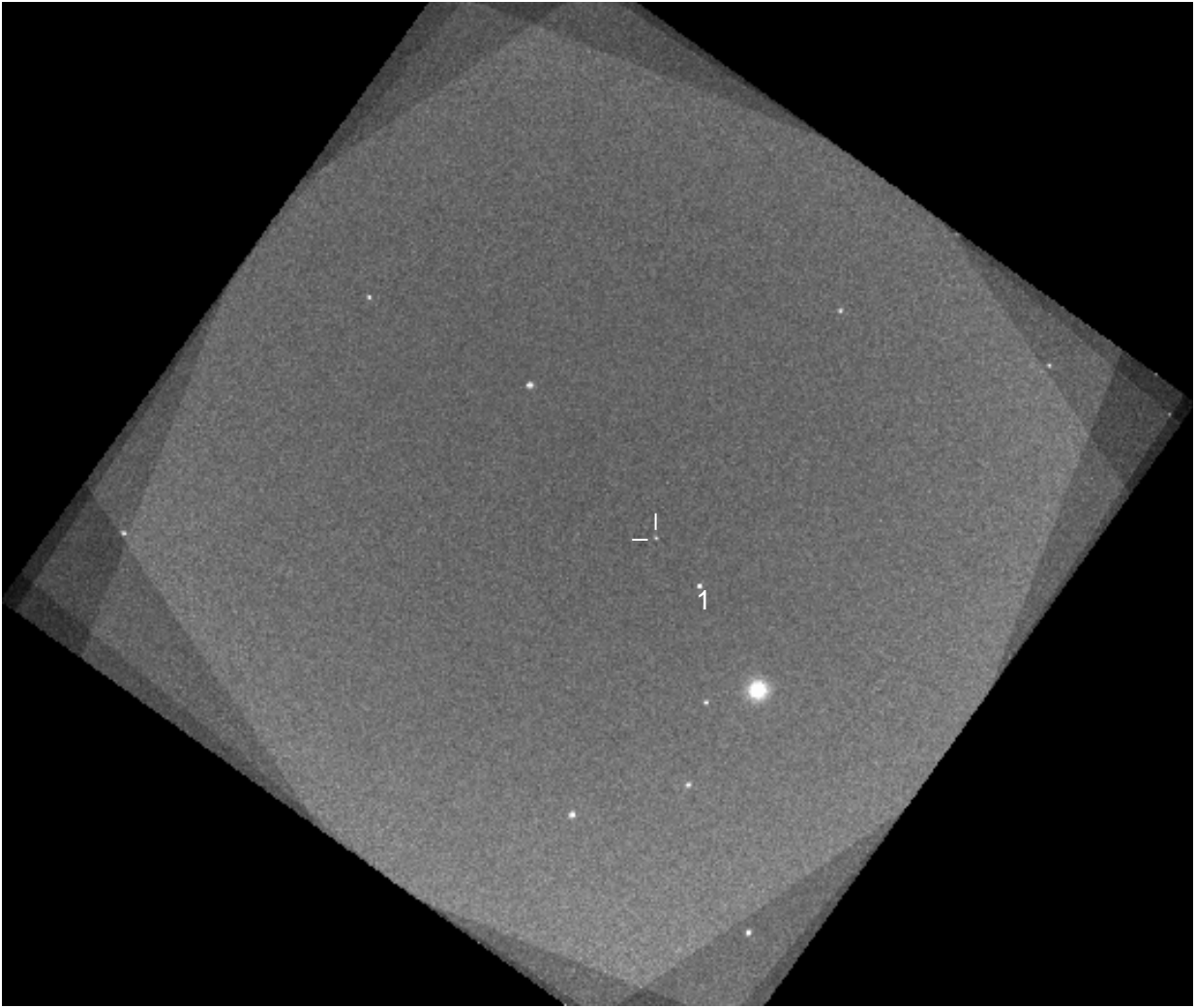} 
\caption{STIS/NUV-MAMA image of the field of PSR 1929+10 taken through
the  F25QTZ filter (epoch  2001.73). The  image is  the result  of the
combination of 10 exposures taken at five different epochs for a total
exposure time of 12\,000 s (see  text).  The frame is aligned in Right
Ascension and Declination  (North to the top, East  to the left).  The
difference  in  the  exposure map  across  the  field  is due  to  the
coaddition  of images taken  with different  roll angles.   The pulsar
candidate counterpart is marked by  the two ticks.  As a reference, we
have labelled star 1 from Figure 2c of Pavlov et al.  (1996)
\label{fig1}}
\end{figure}

\begin{figure}
\plotone{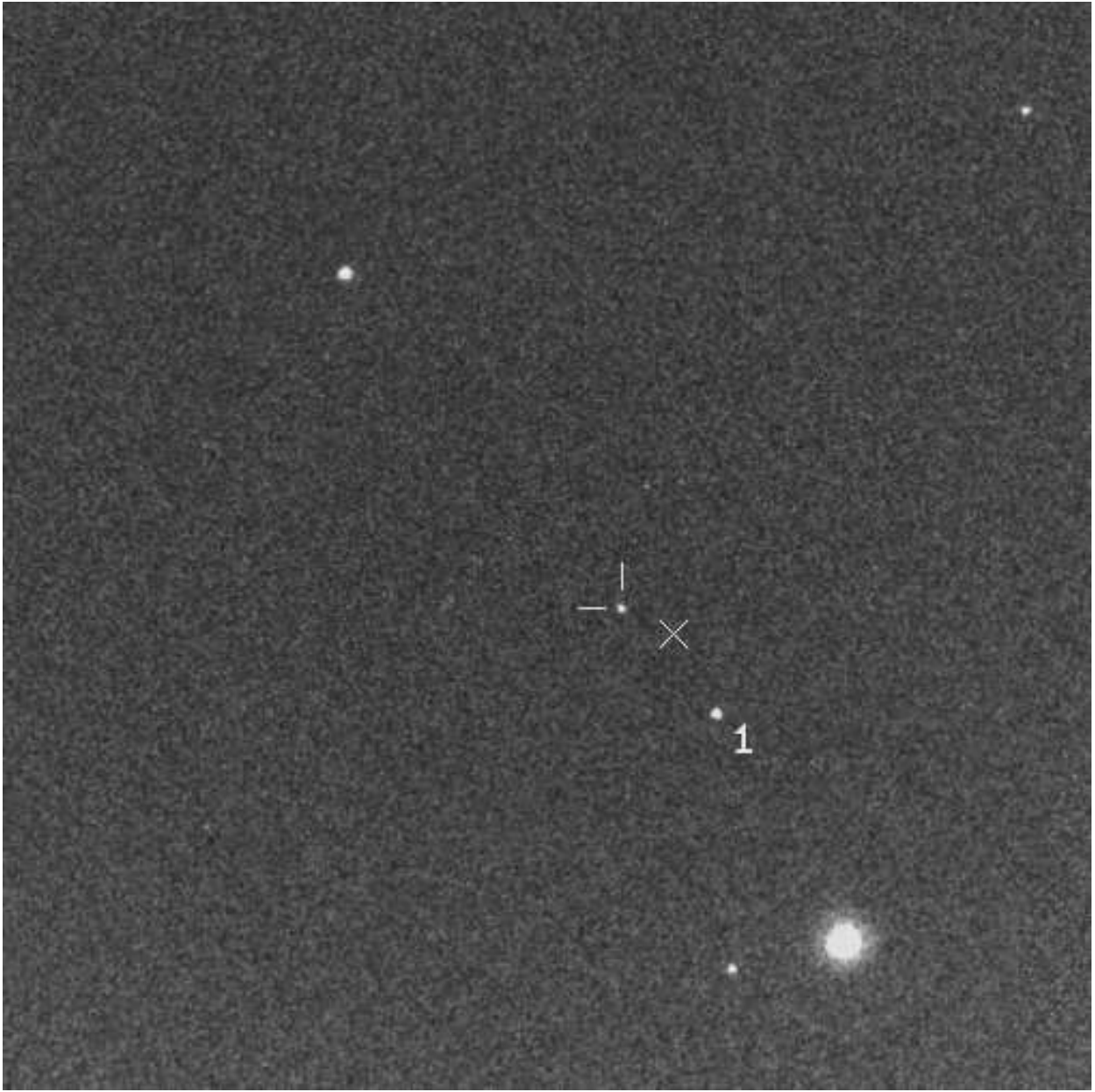} 
\caption{Close-up of Figure 1 centered around the pulsar position.  The cross
marks  the relative  coordinates of  the  pulsar at  epoch  1994.52,
corresponding to the FOC observations of  Pavlov et al. (1996). The pulsar
displacement in 7.2 years is evident.  \label{fig2}}
\end{figure}

\begin{figure}
\plotone{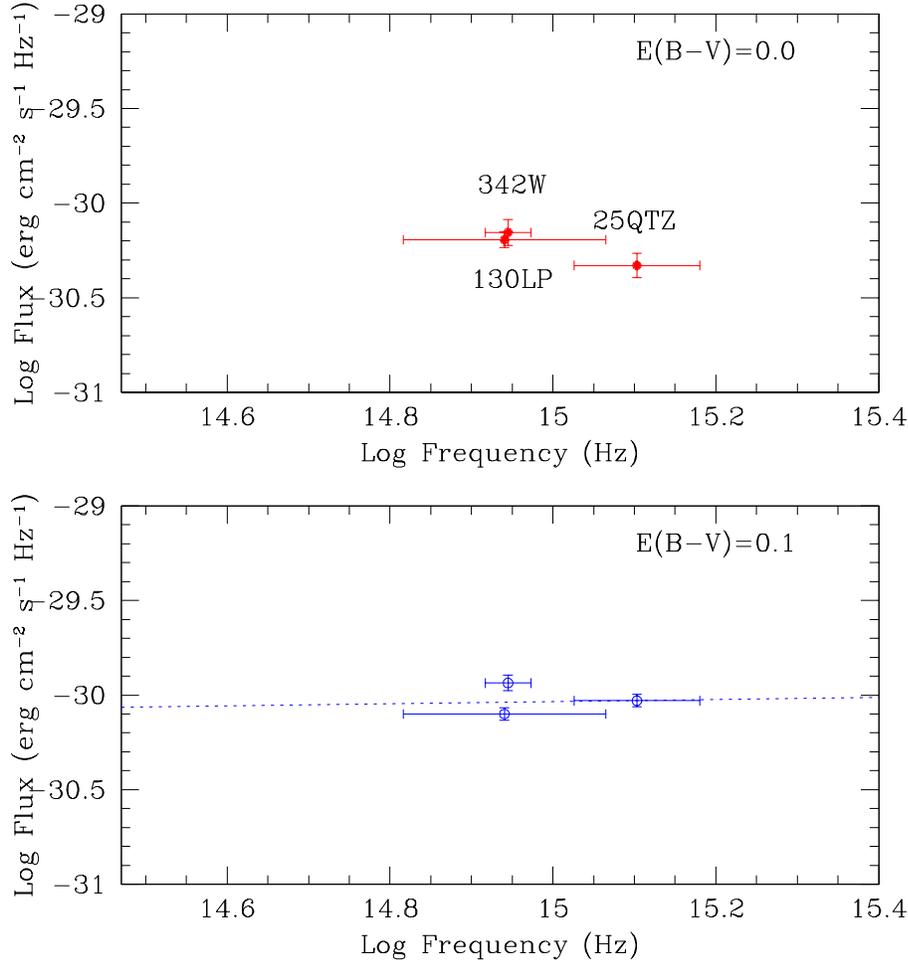} 
\caption{Flux of the PSR 1929+10  counterpart in the STIS 25QTZ filter
compared with  the FOC  measurements in the  130LP and  342W passbands
(Pavlov  et  al. 1996).  In  the upper  panel  no  correction for  the
interstellar  extinction has  been applied.  In the  lower  panel, the
fluxes  have   been  corrected  for  an   interstellar  extinction  of
$E(B-V)=0.1$. In both cases, the  dashed line represents the power law
best fitting the spectral data.  \label{fig3}}
\end{figure}

\end{document}